# OESPA: A Theory of Programming that Support Software Engineering

Chongyi Yuan
School Of Electronics Engineering
And Computer Science
Peking University
Beijing, China
lwyuan@pku.edu.cn

Wen Zhao, Sen Ma
National Engineering Research Center for Software Engineering
Peking University
Beijing, China
zhaowen@pku.edu.cn , masen@pku.edu.cn

**Abstract**

A new theory of programming is proposed. The theory consists of OE (Operation Expression), SP (Semantic Predicate) and A (Axiom), abbreviated as OESPA.

OE is for programming: its syntax is given by BNF formulas and its semantics is defined by axioms on these formulas. Similar to predicates in logic, SP is for describing properties of OE (i.e. programs) and for program property analysis. But SP is different from predicates, it directly relates the final values of variables upon termination of a given OE with initial values of these variables before the same OE. As such, it is feasible to prove or disprove whether a given SP is a property of a given OE by computation based on A (Axioms). SP calculus is proposed for program specification and specification analysis, that is missing in software engineering.

## 1. INTRODUCTION

*1.1 Software engineering need a foundation*

Mathematics, physics, dynamics etc. provide a firm foundation for civil engineering. As such, a blueprint can be approved, prior to its construction, by computing expectable properties against user's requirements.

Is there a way to compute program properties from given program text? Or, in other words, does software engineering have a foundation comparable to the foundation of civil engineering?

The answer is NO! Not yet for today. The fact is, software products rely on post-development testing for its approval. Formal methods do not help much in this regard. For example, model checking requires a formal model to be built from a given program in order to prove properties of it. The consequence is, properties proven by model checking are not necessarily properties of the program[1,2]. They are just the properties of the formal model.

Software engineering needs a firm foundation so that properties of its products can be computed directly from the products.

*1.2 An investigation: formal semantics is missing from programming languages in use*

A programming language has a formal syntax for its compiler, but no formal semantics for property analysis. To be more precise, the BNF formulas that define a programming language have no concern about formal semantics.

For example, the two appearances of $x$ in "$x := x+1$" would be recognized by the compiler as a unique identifier, representing the same variable. But a single variable is a two-facet semantic object: it represents a memory location as well as a data stored in that location.

The way in which variables are used by programming languages has obscured this semantic difference. This is why a assignment like "$x := x+1$" has always been treated as a whole in the discipline of formal semantics, by saying that "the semantics of $x := x+1$" is "an operation", "a function", or "a denotation". In Hoare logic[3,4], the same semantics is given by assertions

$$\forall a \in D : \{x = a\} \, x := x+1 \, \{x = a+1\}$$

where $D$ is the domain of $x$.

This observation explains why program properties can not be computed from program text. Formal semantics is missing in the BNF formulas that define programming languages in use.

*1.3 Predicate and program semantics*

Predicates are suitable for formal description of state properties of programs. But, state properties are not the only thing important to program semantics. An even more important aspect of program semantics is how two consecutive states are related with each other.

For example, let $S$ be a program (or program segment) to exchange values of variables $x$ and $y$. As long as $x$ and $y$ hold valid data, $S$ should do the exchange, regardless what exact values are there in $x$ and $y$. With the predicates as the means, this property has to be given indirectly by

$$\forall a,b \in D : \{x = a \wedge y = b\} \, S \, \{x = b \wedge y = a\} \quad (1)$$

where $D$ is the domain of both $x$ and $y$.

A straightforward description of this property would be

$$x' = y \wedge y' = x \quad (2)$$

where $x', y'$ represent respectively the final values of $x$ and $y$ upon termination of $S$.[5]

It is clear at a glance that (2) is much simpler and more explicit than (1). The point is, how to relate (2) with a given program? Or, how to prove formally that (2) is a property of $S$ when the text of $S$ is given?

We have proposed a new way of programming in the last few years[6,7,8,9], in which assignments are operations on physical object (one of the two facets of variables) and programs are expressions consisting of such operations and control operators. This new type of expressions is called Operation Expression, OE for short.

OE has a formal syntax as well as a formal semantics, and the former is given by BNF formulas while the latter is given by axioms based on BNF formulas.

Semantic axioms are given in terms of two read_operations. In addition to the conventional instant read_operation on variables, two new read_operations are introduced: to read a variable before or after an OE. What obtained by such a read operation is a mathematical expression, telling how a variable before or after the given OE is related to initial values. To read a variable before an OE is meaningful when that OE is a constituent portion of another OE.

Recall the above mentioned program $S$. If $S$ is given as an OE, then $\underline{x}(S)$ and $\underline{y}(S)$, where "_" is the read_after operator, tell respectively how $x$ and $y$ after $S$ is related to variables before $S$. So, in case

$$\underline{x}(S) = y \wedge \underline{y}(S) = x \quad (3)$$

is true, $S$ does exchange the values of $x$ and $y$. Note that $\underline{x}(S)$ and $\underline{y}(S)$ are computable from $S$ (i.e. OE) based on semantic axioms. Here $S$ is considered a stand-alone program, not part of anther OE.

If $S$ is removed from (3), we get

$$\underline{x} = y \wedge \underline{y} = x \quad (4)$$

(4) is called a semantic predicate, SP for short, $\underline{x}$ and $\underline{y}$ are "functions" defined on OE. For any given OE, say $P$, we can compute $\underline{x}(P)$ and $\underline{y}(P)$ based on semantic axioms on OE, and to check whether $\underline{x}(P) = y \land \underline{y}(P) = x$ is true. (4) is a property of $P$ if and only if $\underline{x}(P) = y \land \underline{y}(P) = x$ is true. It is in this sense, we say that program properties can be computed.

Recall (2) given above, (4) is as direct as (2) and as formal as (2), but we know how to compute $\underline{x}$ and $\underline{y}$ for a given program while there is no formal connection between $x', y'$ and a given program.

*1.4 3-step formalism: how this paper is organized*

A general understanding of formalism may be stated as "to express something in precise and rigorous mathematics". But, this understanding does not bring much hint with it about how to do formalism. Here we proposed a 3-step approach as a guideline for practising formalism. The 3 steps are: abstraction, representation and analysis methods.

- Abstraction

Physical objects should be abstracted as mathematical objects, based on the purpose of the formalism in question, by forgetting physical aspects that have on connection with the purpose.

- Representation

This step focuses on how mathematical objects obtained by abstraction are related with each other. Such relations usually form a mathematical system like an equation system.

- Analysis methods

To develop and/or to apply analysis methods for the mathematical system to solve problems raised by the purpose of the formalism.

Formalism has focused too much on the representation step so far, with the abstraction step more or less overlooked. The practice of defining programming languages is but one such example. The purpose of our study is to propose a theory of programming that may serve as the foundation of software engineering, and that is comparable to the foundation of civil engineering. To this end, we followed the 3-step approach in practice.

- Abstraction

Programming languages in use have abstracted memory locations as mathematical variables. A mathematical variable allows a read operation applied on it: whenever it appears in a mathematical expression, the read operation is implied. This is an instant read (the returned value is always the value currently held by the variable) and needs no operator. To cope with the need of programming, assignments on the variables are introduced with ":=" as the assign operator for most of programming languages. But, as an operator ":=" is not well defined since its semantics is not formally given. That means, the abstraction is incomplete.

We propose a complete abstraction of memory locations as a program variable in section 2 as the basis for BNF formulas that define the syntax of OE. Section 2 is about BNF formulas.

- Representation

A mathematical expression relates mathematical variables with each other. An OE (Operation Expression) defined by BNF formulas given in section 2 relates program variables with each other. Thus, what an OE provides is a syntactic representation.

All mathematical operators have well defined semantics so that every mathematical expression can be computed to yield a value. We defined semantics for OE operators with semantic axioms in section 3, so that every OE can be executed. Thus, semantic axioms provide a semantic representation of how program variables are related.

This paper focuses on sequential operation expressions, SOE for short, while parallel operation expressions (POE) and reactive operation expressions (ROE) are left for separate discussion. Operators for

SOE include the write_operator (i.e. assign operator), and control operator to form operation sequences, choices and loops[10].

- Analysis methods

Section 4 is about theorems derived from semantic axioms, Section 5 discusses properties of SOE and Section 6 proposes a SP calculus for program specification and specification analysis. These 3 sections are about methods for SOE property analysis. Section 7 relates SP calculus with SOE specification, Section 8 points out how to derive SOE from specification, and section 9 talks about SOE and software engineering: why OESPA may serve as the foundation of software engineering. Section 10, the last section, plans our future work and expresses our thanks to those who have supported our research. References are given to end this paper.

## 2. BNF FORMULAS FOR OE SYNTAX

### 2.1 Where to start

We don't start from scratch. Data types, scalar or structured, are well defined, syntactically and semantically, and shared by programming languages. OE assumes their definitions without redefining them. It is in this sense, OE is not a complete programming language.

Furthermore, we don't make it explicit what data types are included in OE. It is left for implementation of OE to decide when it becomes mature enough for practical use. For structured data types, array is used as an example to show how to be included in OE. OE is open to everything: as long as it has a well-defined syntax and a well-defined semantics that fit in OE, it may be included, in a nature way, as a constituent of OE.

In the rest of this paper, the word "variable" will mean a scalar variable, an array element or an array as a whole, unless otherwise stated. The range of all arrays are assumed to be $[0..N]$.

### 2.2 Preliminary

#### 2.2.1 Operations on variables

Let $V$ be the set of variables in question.

There are 3 operations on a variable, namely the write_operation, the read_after_operation and read_before_operation. The over_bar " ¯ ", the under_bar " _ ", and the curved under_bar " $\sim$ " are respectively the operator as given below:

$\bar{\phantom{x}} : V \times E_M \to E_0$, i.e. for $x \in V$ and $e \in E_M$, $\bar{x}(e) \in E_0$,

$\underline{\phantom{x}} : V \times E_0 \to E_M$, i.e. for $x \in V$ and $p \in E_0$, $\underline{x}(p) \in E_M$,

$\underset{\sim}{\phantom{x}} : V \times E_0 \to E_M$, i.e. for $x \in V$ and $p \in E_0$, $\underset{\sim}{x}(p) \in E_M$,

where $E_M$ is the set of mathematical expression on $V$ (i.e. every expression in $E_M$ contains only variables in $V$), $E_0$ is the set of OE to be defined. These operators are binary: "$\times$" represents Cartesian product.

Let $x \in V$ and $e \in E_M$, then $\bar{x}(e) \in E_0$, and $\underline{x}(\bar{x}(e)) \in E_M$ and $\underset{\sim}{x}(\bar{x}(e)) \in E_M$. For example, let be $e \equiv x+y$, then we have, by intuition, $\underline{x}(\bar{x}(x+y)) = x+y$, and $\underset{\sim}{x}(\bar{x}(x+y)) = x$ when $\bar{x}(x+y)$ is studied in isolation. Semantic axioms specify how these operators are related as suggested by above examples from intuition.

#### 2.2.2 Variable functions and Semantic predicates

From the definitions of the read_after_operation "_" and the read_before_operation "$\sim$":

$\underline{\phantom{x}} : V \times E_0 \to E_M$

and
$$\tilde{} : V \times E_0 \to E_M$$

We have, for every $v$ in $V$:
$$\overline{v} : E_0 \to E_M$$
and
$$\underline{v} : E_0 \to E_M$$

**Definition 1**

1. $\overline{v}$ and $\underline{v}$ are called variable functions: $\overline{v}$ is the final function and $\underline{v}$ is the initial function, F_function and I_function for short respectively.

2. $\overline{V} = \{\overline{v} | v \in V\}$, the set of all F_functions, is called the F_set,
$\underline{V} = \{\underline{v} | v \in V\}$, the set of all I_functions, is called the I_set.

3. For operation expression $p$,
$\overline{V}(p) = \{\overline{v}(p) | v \in V\}$, is called the F_set of $p$,
$\underline{V}(p) = \{\underline{v}(p) | v \in V\}$, is called the I_set of $p$.

4. For operation expressions $p$ and $q$,
$\overline{V}(p) = \overline{V}(q)$ if and only if $\forall v \in V : \overline{v}(p) = \overline{v}(q)$,
$\underline{V}(p) = \underline{V}(q)$ if and only if $\forall v \in V : \underline{v}(p) = \underline{v}(q)$,
$\underline{V}(p) = \overline{V}(q)$ if and only if $\forall v \in V : \underline{v}(p) = \overline{v}(q)$.

**Example 1**

Let $p \equiv \overline{y}(x+1)$ and $q \equiv \overline{x}(x+1)$ be operation expressions, where operator "$\overline{\phantom{x}}$" takes $y$ and $x+1$, respectively $x$ and $x+1$ as its first and second operands. Let $S \equiv q; p$ where "$;$" is the sequential control operator shared by programming languages. We have, from intuition,
$$\overline{y}(\overline{y}(x+1)) = x+1 \quad \text{and} \quad \overline{x}(\overline{x}(x+1)) = x+1$$

i.e. $\overline{y}(p) = x+1$ and $\overline{x}(q) = x+1$

In the context of $S$, we have
$$\underline{x}(p) = \overline{x}(q) = x+1$$
and $\overline{y}(S) = \overline{y}(p) | (\underline{x}(p) = \overline{x}(q))$

$$= \underline{x}(p) + 1 = (x+1) + 1 = x + 2$$

where $\overline{y}(p) | (\underline{x}(p) = \overline{x}(q))$ indicates that the initial value of $x$ before $p$ is the final value of $x$ after $q$.

Note from this example, $x$ is the very initial value before $S$. $x+1$ is the final value after $q$ as well as the initial value before $p$. Thus, $x+2$ is the final value after $S$. The exact value of $x$ is irrelevant in above computation.

**Definition 2**

1. A predicate $R$ on $\underline{V} \cup \overline{V}$ is called a semantic predicate if it contains at least one variable function from $\underline{V}$;

2. A Boolean expression on $\underline{V} \cup \overline{V}$ is called a semantic Boolean expression if it contains at least one variable function from $\underline{V}$.

A semantic predicate is denoted by $R(\underline{V}, \overline{V})$ and a semantic Boolean expression is denoted by $\underline{b}$. The domain of semantic predicates is $E_0$ since $\underline{V}$ is the F_set defined on $E_0$.

**Example 2**

$R_1(\underline{V}, \overline{V}) \equiv \underline{x} = y \wedge \underline{y} = x$ , $R_2(\underline{V}, \overline{V}) \equiv \underline{x} \leq \underline{y}$ are semantic predicates, $\underline{x} \leq \underline{y}$, $\underline{x} + x = y$ are semantic Boolean expressions.

$R \equiv x + y + z = 0$ is not a semantic predicate since it contains no variable functions. $x < y$ is a Boolean expression, but not a semantic one since it has nothing to do with $\underline{V}$.

*2.2.3 Conditional expressions*

Let $e_1, e_2, ..., e_n$ be expressions of the same type and $b_1, b_2, ..., b_n$ be Boolean expressions, where $n$ is an integer with $n > 1$.

**Definition 3**

1. $e_1$ if $b_1 \sim e_2$ if $b_2 \sim ... \sim e_n$ if $b_n$ is called a conditional expression, if

$$\forall i, j : 1 \leq i, j \leq n : (i \neq j \wedge b_i \wedge b_j \to e_i = e_j)$$

$$\wedge (b_1 \vee b_2 \vee ... \vee b_n)$$

2. Let $l$ be the value of the above defined conditional expression, then $\forall i : 1 \leq i \leq n : (b_i \to l = e_i)$

This definition ensures that every conditional expression has a unique value if every $e_i, i = 1, 2, ..., n$, has a value at a given state.

A conditional expression is also called a table by some formal methods like PVS, where $\forall i, j : 1 \leq i, j \leq n : (i \neq j \to \neg(b_i \wedge b_j))$ is required. We have this requirement relaxed a bit here.

Conditional expressions are mainly used in semantic axioms.

*2.3 BNF Formulas*

In the BNF formulas given below, $v$ represents either a scalar variable like $x$, or an array element like $A[i]$, or an array as a whole like $A$. All arrays are assumed to share $[0..N]$ as their index range. In addition, type match is always assumed whenever it is required.

Conventions we follow are:

$i, j$: variables for array index,

$n, N$: integer constants, $n > 0$, $N > 1$,

$E$: array expressions,

$\underline{b}$: semantic Boolean expressions,

$b, b_1, b_2, ..., b_n, b(i)$: Boolean expressions,

$e, e_1, e_2, ..., e_n, e(i)$: mathematical expressions

BNF formulas

$(B1) = (B1.0) + (B1.1) + (B1.2) + (B1.3)$

$(B1.0)$ simple_term ::=<s_term> | <vs_term> | <es_term>

$(B1.1)$ s_term ::= $\overline{x}(e) | \overline{A}[i](e) | \overline{A}(E)$

$(B1.2)$ vs_term ::= $\varepsilon | \overline{v}(e)^{\underline{b}}$

$(B1.3)$ es_term ::= $\overline{v}(e_1)^{\underline{b_1}} \overline{v}(e_2)^{\underline{b_2}} ... \overline{v}(e_n)^{\underline{b_n}}$

Remarks

• A simple_term applies write_operation on a single variable.

• A vs_term is a variation of s_term: it is either the special term $\varepsilon$, or a conditional term $\overline{v}(e)^{\underline{b}}$.

• An es_term is an extension of a s_term: a multiple choices for the same variable. It is required by semantic axioms to be given that

$$i \neq j \wedge b_i \wedge b_j \to e_i = e_j$$

$(B2) = (B2.0) + (B2.1) + (B2.2) + (B2.3)$

$(B2.0)$ multi_term ::=<m_term> | <vm_term> | <am_term>

$(B2.1)$ m_term ::=simple_term • simple_term | m_term • simple_term

$(B2.2)$ vm_term ::=m_term$^{\underline{b}}$

$(B2.3)$ am_term ::= $\overline{A}[i : b(i)](e(i))$

Remarks

• A multi_term applies write_operation on more than one variable.

• A vm_term is a conditional m_term.

• An am_term is a conditional m_term for elements of the same array.

• The operator "•" between two simple_terms is often omitted, just like the operator for multiplication.

$(B3) = (B3.0) + (B3.1) + (B3.2) + (B3.3)$

$(B3.0)$ sequential_O_expression::=<soe> | <v_soe> | <r_soe> | sequential_O_expression;sequential_O_expression

$(B3.1)$ soe::=<term> | <term>;soe

term::=simple_term | multi_term

(B3.2)  v_soe::=soe$^b$

(B3.3)  r_soe::=soe$^n$ | soe$^{\underline{b}}$

Remarks
- ";" is the sequential control operator
- r_soe is for loops
- SOE is used as an abbreviation of Sequential_O_expressions, and sometimes it is also used to denote the set of all Sequential_O_expressions when no confusion possible. Thus, SOE is a subset of $E_0$.

## 3. SEMANTIC AXIOMS ON SOE

Let $p, q$ be SOE, i.e. $p, q \in$ SOE, and let $V_p = \{v \mid \overline{v}(e) \in p\}$ be the subset of $V$, consisting of variables on which $p$ applies write_operation.

(A1) = (A1.1) + (A1.2) + (A1.3), Foundation

(A1.1)  $p = q \Leftrightarrow \forall v \in V : \underline{v}(p) = \underline{v}(q)$

(A1.2)  $v \notin V_p \rightarrow \underline{v}(p) = v$

(A1.3)  $v \in V \rightarrow \underline{v}(p) = v$  if $p$ is in isolation.

(A2) = (A2.1) + (A2.2) + (A2.3) + (A2.4)  simple_term

(A2.1)  $\underline{v}(\overline{v}(e)) = e$

(A2.2)  $V_\varepsilon = \phi$

(A2.3)  $\overline{v}(e)^b = \overline{v}(e)$ if $b \sim \varepsilon$ if $\neg b$

(A2.4)  $\overline{v}(e_1)^{b_1} \overline{v}(e_2)^{b_2} ... \overline{v}(e_n)^{b_n} =$
$\overline{v}(e_1)$ if $b_1 \sim \overline{v}(e_2)$ if $b_2 \sim ... \sim \overline{v}(e_n)$ if $b_n$
$\sim \varepsilon$ if $\neg(b_1 \vee b_2 \vee ... \vee b_n)$

Distributive law on conditional operation expressions:
$\underline{v}(p_1$ if $b_1 \sim p_2$ if $b_2 \sim ... \sim p_n$ if $b_n)$
$= \underline{v}(p_1)$ if $b_1 \sim \underline{v}(p_2)$ if $b_2 \sim ... \sim \underline{v}(p_n)$ if $b_n$

(A3) = (A3.1) + (A3.2) + (A3.3)  multi_term

(A3.1)  $\underline{v}(p \bullet q) = \underline{v}(p)$ if $v \in V_p \sim \underline{v}(q)$ if $v \in V_q$  for $v \in V_{p \bullet q}$

(A3.2)  $(p \bullet q)^b = p^b \bullet q^b$

(A3.3)  $\overline{A}[i : b(i)](e(i))$
$= \overline{A}[0](e(0))^{b(0)} \bullet \overline{A}[1](e(1))^{b(1)} \bullet ... \bullet \overline{A}[N](e(N))^{b(N)}$

(A4) = (A4.1) + (A4.2) + (A4.3) + (A4.4) + (A4.5) + (A4.6)
$p, q, p_1, p_2, p_3 \in$ SOE

(A4.1)  $\underline{v}(p; q) = (\underline{v}(q) \mid \underline{V}(q) = \underline{V}(p))$  for all $v \in V$

(A4.2)  $p_1; p_2; p_3 = (p_1; p_2); p_3$

(A4.3)  $p^b = p$ if $b \sim \varepsilon$ if $\neg b$

(A4.4)  $p^n = p$ if $n = 1 \sim p^{n-1}; p$ if $n > 1$

(A4.5)  $p^{\underline{b}} = \varepsilon$ if $\underline{b}(\varepsilon) \sim p^n$ if $\neg \underline{b}(\varepsilon)$
$\wedge \exists n : (\underline{b}(p^n) \wedge \forall l : 0 < l < n : \neg \underline{b}(p^l)) \sim p; p^{\underline{b}}$  otherwise

(A4.6)  $\underline{v}(p; q) = \underline{v}(p)$  for all $v \in V$

Remarks
- $\underline{v}(q) \mid \underline{V}(q) = \underline{V}(p)$ indicates that $q$ takes $\underline{u}(p)$ as the initial value of $u$ for every $u \in V$.
- $p; p^{\underline{b}}$ in (A4.5) represents an endless repetition of $p$. For SOE at the mean time, this is a semantic error. We keep it as it is since we will extend SOE to include reactive actions.

## 4. THEOREMS DERIVED FROM SEMANTIC AXIOMS

**Theorem 1**
All axioms given with conditional expressions, namely (A2.3),(A2.4),(A3.1),(A4.3) and (A4.4),(A4.5) are deterministic.

**Proof.**
Conclusions in this theorem are ensured by (A1.1). All those that do not fulfill this theorem, are not semantically valid, and should be excluded from SOE.

**Definition 4**

A sequential operation expression $p$, syntactically satisfies all BNF formulas given in subsection 2.3, is a valid SOE if and only if it is semantically deterministic, i.e. it leads to no ambiguity no matter which semantic axiom is applied.

In what follows SOE refers to only valid sequential operation expressions.

**Theorem 2**

$(p_1; p_2); p_3 = p_1;(p_2; p_3)$ for $p_1, p_2, p_3 \in SOE$

**Proof.**

$\underline{V}((p_1; p_2); p_3) = \underline{V}(p_3) | \underline{V}(p_3) = \underline{V}(p_1; p_2)$ (A4.1)
$= \underline{V}(p_3) | \underline{V}(p_3) = (\underline{V}(p_2) | \underline{V}(p_2) = \underline{V}(p_1))$ (A4.1)
$= (\underline{V}(p_3) | \underline{V}(p_3) = \underline{V}(p_2)) | \underline{V}(p_2; p_3) = V(p_1)$ (A4.6)
$= \underline{V}(p_2; p_2) | \underline{V}(p_2; p_3) | \underline{V}(p_1)$ (A4.1)
$= \underline{V}(p_1;(p_2; p_3))$ (A4.1)
i.e. $(p_1; p_2); p_3 = p_1;(p_2; p_3)$ (A1.1)

**Theorem 3**

$(p^{b_1})^{b_2} = p^{b_1 \wedge b_2}$ (A4.3)

**Corollary 1** for terms
1. $\varepsilon \bullet p = p \bullet \varepsilon = p$ (A3.1)
2. $p \bullet p = p$ (A3.1)
3. $p \bullet q = q \bullet p$ (A3.1)
4. $(p_1 \bullet p_2) \bullet p_3 = p_1 \bullet (p_2 \bullet p_3)$ (A3.1)

## 5. PROPERTIES OF SOE

**Definition 5**

$R(\underline{V}, V)$ is a property of $p$, $p \in SOE$, if $R(\underline{V}(p), V)$, i.e. $p$ makes $R(\underline{V}, V)$ true.

**Example 3**

1. $\underline{x} = y$ is a property of $\overline{x}(y)$ since $\underline{x}(\overline{x}(y)) = y$ by (A2.1).

2. $\underline{x} \le y$ is a property of $\overline{x}(y)^{x>y}$ since, by (A2.3),

   $\underline{x}(\overline{x}(y)^{x>y}) = y$ if $x > y$ ~ $x$ if $x \le y$.

3. $\underline{x} = y \wedge \underline{y} = x$ is a property of

   $p \equiv \overline{x}(x+y); \overline{y}(x-y); \overline{x}(x-y)$

Detailed Computations for 2 and 3:

2. $p \equiv \overline{x}(y)^{x>y}$,

   $p = \overline{x}(y)$ if $x > y$ ~ $\varepsilon$ if $x \le y$ (A2.3)

   $\underline{x}(p) = \underline{x}(\overline{x}(y))$ if $x > y$ ~ $\varepsilon$ if $x \le y$)

   $= \underline{x}(\overline{x}(y))$ if $x > y$ ~ $\underline{x}(\varepsilon)$ if $x \le y$

   (by distributive law)
   $= y$ if $x > y$ ~ $x$ if $x \le y$ (A2.1),(A2.2),(A1.2)
   i.e. $\underline{x}(p) \le y$

3. $p \equiv \overline{x}(x+y); \overline{y}(x-y); \overline{x}(x-y)$

   $= (\overline{x}(x+y); \overline{y}(x-y)); \overline{x}(x-y)$ (A4.2)

   $\underline{x}(\overline{x}(x+y)) = x+y, \underline{y}(\overline{x}(x+y)) = y$ (A2.1),(A1.2)

   so,

   $\underline{x}(\overline{x}(x+y); \overline{y}(x-y))$

   $= (\underline{x}(\overline{y}(x-y)) | \underline{x} = x+y, \underline{y} = y)$ (A4.1)

   $= x + y$

   $\underline{y}(\overline{x}(x+y); \overline{y}(x-y)) = (\underline{y}(\overline{y}(x-y)) | \underline{x} = x+y, \underline{y} = y)$

   $= (x+y) - y = x$

   Thus,

   $\underline{x}(p) = \underline{x}(\overline{x}(x-y)) | \underline{x} = x+y, \underline{y} = x$

   $= (x+y) - x = y$

   $\underline{y}(p) = \underline{y}(\overline{x}(x-y)) | \underline{x} = x+y, \underline{y} = x$

$= x$

**Definition 6**

Semantic predicate $R(\underline{V},V)$ is bipartite if it takes the form $R(\underline{V},V) \equiv R'(\underline{V})\ op\ R'(V)$, where $op \in \{=,<,\leq,>,\geq,\rightarrow,\leftarrow\}$ and $R'(\underline{V})$, $R'(V)$ share the same structure with variables from $\underline{V}$ and $V$ respectively.

Note that $op$ is a binary transitive operator, "$\rightarrow$" is the "implies" operator and "$\leftarrow$" is the "implied by" operator.

**Example 4**

$R_1(\underline{V},V) \equiv \underline{x}+\underline{y}+\underline{z} = x+y+z$, $R_2(\underline{V},V) \equiv \underline{x} < x$

and $R_3(\underline{V},V) \equiv \underline{x}+\underline{y} > 0 \leftarrow x+y > 0$ are bipartite.

$R_4(\underline{V},V) \equiv \underline{x} < \underline{y} \leftarrow x < z$ is not bipartite.

**Definition 7**

If bipartite $R(\underline{V},V) \equiv R'(\underline{V})\ op\ R'(V)$ is a property of $p$, then $R'(\underline{V})$ is said to be, as shown by the table below, stable, inheritable, traceable, respectively, constant, decreasing, not increasing, increasing and not decreasing. Such $R'(\underline{V})$ will be denoted by, respectively, stable($R'(\underline{V})$), inherit($R'(\underline{V})$), traceable($R'(\underline{V})$), const($R'(\underline{V})$), dec($R'(\underline{V})$), not_inc($R'(\underline{V})$), inc($R'(\underline{V})$) and not_dec($R'(\underline{V})$).

| $op$ | $R'(\underline{V}) \in E_L$ | $op$ | $R'(\underline{V}) \notin E_L$ |
|---|---|---|---|
| $=$ | stable | $=$ | constant |
| $\leftarrow$ | inheritable | $<$ | decreasing |
| $\rightarrow$ | traceable | $\leq$ | not increasing |
| $E_L$: Set of logic expression | | $>$ | increasing |
| | | $\geq$ | not decreasing |

Table 1  $R'(\underline{V})\ op\ R'(V)$ is a property of $p$

**Theorem 4**

If $R(\underline{V},V) \equiv R'(\underline{V})\ op\ R'(V)$ is a property of $p$, then

1. stable($R'(\underline{V})$) $\land\ R'(V) \rightarrow R'(\underline{V}(p))$
2. inherit($R'(\underline{V})$) $\land\ R'(V) \rightarrow R'(\underline{V}(p))$
3. traceable($R'(\underline{V})$) $\land\ \neg R'(V) \rightarrow \neg R'(\underline{V}(p))$

The proof is straightforward from Definition 7.

**Definition 8**

A semantic predicate is a state predicate if it contains only variable functions. A state predicate $R$ is denoted by $R(\underline{V})$.

**Example 5**

$R_1(\underline{V}) \equiv \underline{x}+\underline{y}+\underline{z} = 100$, $R_2(\underline{V}) = \underline{x} \leq \underline{y} \land \underline{y} \leq \underline{z}$ are state predicates.

**Definition 9**

1. $R(\underline{V})$ is an invariant of $p$, if $R(\underline{V}(\varepsilon)) \land R(\underline{V}(p))$.
2. $R(\underline{V})$ is a conditional property of $p$ if there exists a state predicate $Q(\underline{V})$ such that $Q(\underline{V}(\varepsilon)) \rightarrow R(\underline{V}(p))$. $Q(\underline{V})$ is called the condition of $(R(\underline{V}),p)$.

**Example 6**

1. Let be $p \equiv \overline{x}(x+1)\overline{y}(y-1)$, $R_1(\underline{V}) \equiv \underline{x}+\underline{y}+\underline{z} = 100$ and $R_2(\underline{V}) \equiv \underline{x}+\underline{y}+\underline{z}$, then $R_1(\underline{V})$ is an invariant of $p$ if $x+y+z = 100$ initially, and $R_2(\underline{V})$ is a constant of $p$ regardless $x+y+z = 100$ or not. Note $R_2(\underline{V})$ is itself not a predicate.

2. $Q(\underline{V}) \equiv \underline{x} \leq \underline{y} \land \underline{x} \leq \underline{z}$ is the condition for $((\overline{y}(z)\overline{z}(y))^{y>z}, \underline{x} \leq \underline{y} \land \underline{y} \leq \underline{z})$.

**Definition 10**

State predicate $R(\underline{V})$ is irrelevant to $p$, $p \in SOE$,

if and only if $V_R \cap V_p = \phi$ where $V_R = \{v \mid \underline{v} \in R(\underline{V})\}$ and $V_p = \{v \mid \overline{v(e)} \in p\}$.

**Example 7**

$R(\underline{V}) \equiv \underline{x} \leq \underline{y}$ is irrelevant to $p = \overline{z}(x+y)$, since $V_R = \{x, y\}$ and $V_p = \{z\}$.

**Corollary 2**

If $R(\underline{V})$ is irrelevant to $p$, then $v \in V_R \to \underline{v}(p) = v$ and thus $R(\underline{V}(p)) = R(V)$, where $R(V)$ is obtained for $R(\underline{V})$ by replacing every variable function $\underline{v}$ with the variable $v$.

**Example 8**

For $R(\underline{V}) \equiv \underline{x} \leq \underline{y}$ and $p = \overline{z}(x+y)$, we have $\underline{x}(p) = x$ and $\underline{y}(p) = y$. Thus, $R(\underline{V}(p)) \equiv \underline{x}(p) \leq \underline{y}(p)$, i.e. $R(\underline{V}(p)) \equiv x \leq y$.

**Theorem 5**

1. If $R(\underline{V})$ is a conditional property of $p$ with $Q(\underline{V})$ as the condition of $(R(\underline{V}), p)$, $R(\underline{V})$ is irrelevant to $q$ and $Q(\underline{V})$ is a property of $q$, then $R(\underline{V})$ is a property of $q; p$.

2. If $R(\underline{V})$ is a property of $q$ and it is irrelevant to $p$, then $R(\underline{V})$ is a property of $q; p$.

3. If $R(\underline{V})$ is an invariant of both $q$ and $p$, then $R(\underline{V})$ is a property of $q; p$, or to be more precise, $R(\underline{V})$ is also an invariant of $q; p$.

4. For $v \in V_p$, if $\underline{v}(p)$ contains no variable in $V_q$ then $\underline{v}(q; p) = \underline{v}(p)$.

**Proof.**

1, 2 and 3 are direct consequences of relevant definitions.

For conclusion 4,

$\underline{v}(q; p) = \underline{v}(p) | \underline{V}(p) = \underline{V}(q)$ by (A4.1)

For variable $x$ that appears in $\underline{v}(p)$, we know $x \notin V_q$ thus $\underline{x}(q) = x$, i.e. $\underline{x}(p) = x$. Since $x$ is an arbitrary variable appearing in $\underline{v}(p)$, we know

$(\underline{v}(p) | \underline{V}(p) = \underline{V}(q)) = \underline{v}(p)$

so, $\underline{v}(q; p) = \underline{v}(p)$.

**Theorem 6**

1. If $R(\underline{V})$ is an invariant of $p$, then $R(\underline{V})$ is also an invariant of $p^n$ for $n > 1$.

2. If there exists $n, n \geq 0$, such that $p^{\underline{b}} = p^n$ by (A4.5), then $\underline{b}$ is a property of $p^{\underline{b}}$, and every invariant of $p$ is also invariant of $p^{\underline{b}}$.

**Proof.** Direct consequence of Theorem 5.

**Example 9**

Let $p = \overline{i}(i+1)\overline{m}(m + A[i])$, then $R(\underline{V}) \equiv \underline{m} = \sum_{l=0}^{i-1} \underline{A}[l]$ is an invariant of $p^n$ for $n$, $0 < n \leq N$, if initially $i = 1 \wedge m = A[0]$. It's easy to compute that $\underline{i}(p^N) = N+1$ and $\underline{m}(p^N) = \sum_{l=0}^{N} A[l]$, i.e. $q; p^N$ is a program to compute the sum of array elements for array $A$, where $q = \overline{i}(1)\overline{m}(A[0])$.

Note that array $A$ is not changed by $p$, i.e. for any index $l$, $\underline{A}[l](p) = A[l]$ (or $\underline{A}(p)[l] = A[l]$). As such $R(\underline{V})$ can be written as $R'(\underline{V}, V) \equiv \underline{m} = \sum_{l=0}^{i-1} A[l]$, and say $R'(\underline{V}, V)$ is an invariant of $p$ and $p^N$. The difference between $R(\underline{V})$ and $R'(\underline{V}, V)$ in this example is that the former is a state predicate and the latter is not. From now on we relax the definition of an invariant to include semantic predicates like $R'(\underline{V},V)$.

**Example 10**

Let $p$, $q$ and $R'(\underline{V},V)$ as given in example **9**. By theorem 6 we know $R'(\underline{V},V)$ is also an invariant of $p^b$ where $\underline{b} \equiv \underline{i} = N+1$, and $\underline{m}(p^b) = \sum_{l=0}^{N} A[l]$, as long as $i = 1 \wedge m = A[0]$ initially, i.e. $\underline{i}(p^b) = 1 \wedge \underline{m}(p^b) = A[0]$.

So, $q; p^b$ is also a program to compute the sum of all elements of array $A$.

## 6. SP CALCULUS

SP calculus is developed mainly for program specification and specification analysis based on semantic axiom (A 1.2), (A 4.1) and (A 4.6).

A noticeable fact is, the write_operation (i.e assignment) is the only operation that changes a program state. Thus for a complete program specification, there must be an expression $e_v$ for an arbitrary variable $v$ such that $\underline{v} = e_v$ is derivable from it, where $e_v$ contains only variables in $V$, no variable function from $\underline{V}$. SP calculus tries to capture the process of deriving $\underline{v} = e_v$ for all $v$ in $V$, from a given specification. A specification is called a SP formula in this calculus.

### 6.1 Semantic Predicates (SP)

The concept of semantic predicates was defined earlier by Definition 2. But, axiom (A 4.1) indicates that I_functions $\underline{v}$ for $v \in V$ are needed in the process of reasoning. Thus, we will extend the definition to serve the need of SP calculus.

**Definition 11**

A predicate $R$ defined on $V \cup \underline{V} \cup \tilde{V}$ is a semantic predicate, SP for short, if it contains no contradiction.

Apparently, all semantic predicates satisfying Definition 2 remain to be semantic predicates by Definition 11. All semantic predicates takes SOE as their domain since $\underline{V}$, $\tilde{V}$ are respectively the F_set and I_set defined on SOE.

**Example 11**

$\underline{x} < \underline{x}$, $\tilde{x} < x \wedge x < \underline{x}$ are contradictions.

We write $v \in R$ for semantic predicate $R$ if $v$ appears in $R$. Similarly, we write $\underline{v} \in R$ and $\tilde{v} \in R$.

Let be
$V_R = \{v \mid v \in R\}$, $\underline{V}_R = \{\underline{v} \mid \underline{v} \in R\}$ and $\tilde{V}_R = \{\tilde{v} \mid \tilde{v} \in R\}$.

**Definition 12**

$R$ is said to be $V$_free, $\underline{V}$_free and $\tilde{V}$_free respectively, if $V_R = \phi$, $\underline{V}_R = \phi$ or $\tilde{V}_R = \phi$.

$R$ is said to be $V$_complete, $\underline{V}$_complete and $\tilde{V}$_complete respectively, if $V_R = V$, $\underline{V}_R = \underline{V}$ and $\tilde{V}_R = \tilde{V}$.

### 6.2 SP formulas and computation Rules

SP is the simplest formula.

**Definition 13**

SP_formula $::= SP \mid SP;$ SP_formula

In what follows, a SP_formula will be called a formula.

For $R$, $R$ is not $\underline{V}$_complete, we have, by Axiom (A 1.2),

**Computation Rule 1: Extension**

$\underline{v} \notin R \to R \equiv (R \wedge \underline{v} = v)$.

We now develop computation Rules for structured formulas, i.e. for $(R_1; R_2)$ and $(R_1; F)$ where $R_1$, $R_2$ are SP while F is itself a formula.

For given $R$, $R$ is a SP without saying, let

$R(\underline{V}, \underline{V})$ denote what is obtained by replacing every appearance of $\underline{v}$ in $R$ with $\underline{v}$ for all $\underline{v} \in \underline{V}$;

$R(V, \underline{V})$ denote what is obtained by replacing every appearance of $v$ in $R$ with $\underline{v}$ for all $v \in V$;

$R(\underline{V}, V)$ denote what is obtained by replacing every appearance of $\underline{v}$ in $R$ with $v$ for all $\underline{v} \in \underline{V}$;

**Example 12**

For $R \equiv \underline{x} = x^2$, we have $R(\underline{V}, \underline{V}) \equiv \underline{x} = x^2$,

$R(V, \underline{V}) \equiv \underline{x} = \underline{x}^2$ and $R(\underline{V}, V) \equiv \underline{x} = x^2$.

**Theorem 7**

$R(\underline{V}, \underline{V})$ and $R(V, \underline{V})$ are both SP if $R$ is $\underline{V}$_free.

**Proof**

$\underline{V}_R = \phi$ ensures that both $R(\underline{V}, \underline{V})$ and $R(V, \underline{V})$ are contradiction free.

Let $R_1$, $R_2$ be both $\underline{V}$_free SP.

**Definition 14**

1. $R_1$ is said to be independent of $R_2$ if $V_{R_1} \cap V_{R_2} = \phi$.

2. If $R_1$ is not independent of $R_2$, then $R'$ is said to be the independent portion of $R_1$ with respect to $R_2$ if $R'$ is the maximum portion of $R_1$ such that $\underline{V}_{R'_1} \cap \underline{V}_{R_2} = \phi$. This fact is denoted by

   $R'_1 = ind(R_1, R_2)$.

**Example 13**

$R_1 \equiv \underline{x} = \sin x \wedge \underline{y} = \cos x$, $R_2 \equiv \underline{x} = \sqrt{x}$, then

$ind(R_1, R_2) \equiv \underline{y} = \cos x$.

**Definition 15**

Let $R_1$, $R_2$ be $\underline{V}$_free. $R_1$ is said to be complete with respect to $R_2$ if $V_{R_1} \supseteq V_{R_2}$ and for every $\underline{v} \in \underline{V}$, an equation $\underline{v} = e_v$ is derivable from $R_1(\underline{V}, \underline{V})$ such that $e_v$ contains only variable from $V$ and $R_2(V, \underline{V}) | R_1(\underline{V}, \underline{V})$ represents a substitution in $R_2(V, \underline{V})$ that every appearance of $\underline{v}$ is replaced by $e_v$ for every $\underline{v} \in \underline{V}$.

**Example 14**

Let be $R_1 \equiv \{\underline{x}, \underline{y}, \underline{z}\} = \{x, y, z\}$,

$R_2 \equiv \{\underline{x}, \underline{y}, \underline{z}\} = \{x, y, z\}$, we have: $V_{R_1} = V_{R_2} = \{\underline{x}, \underline{y}, \underline{z}\}$

and $R_1(\underline{V}, \underline{V}) \equiv \{\underline{x}, \underline{y}, \underline{z}\} = \{x, y, z\}$,

$R_2(V, \underline{V}) \equiv \{\underline{x}, \underline{y}, \underline{z}\} = \{\underline{x}, \underline{y}, \underline{z}\}$ and

$R_2(V, \underline{V}) | R_1(\underline{V}, \underline{V}) \equiv \{\underline{x}, \underline{y}, \underline{z}\} = \{x, y, z\}$. Thus, $R_1$ is complete with respect to $R_2$.

Note that both $R_1$ and $R_2$ require some sorting of $x, y, z$. Thus, $R_1; R_2$ requires one sorting after another, and is itself a sorting. This example explains why we have Definition 15: preparing for a rule applicable to formula $R_1; R_2$. Definition 15 requires only the existence of $e_v$, not the precise expression $e_v$.

**Computation Rule 2: Resolution**

If $R_1$ is complete with respect to $R_2$, then $R_1; R_2 \to R_1' \wedge (R_2(V, \underline{V}) | R_1(\underline{V}, \underline{V}))$ where $R_1' \equiv ind(R_1, R_2)$

Semantic axiom(A 4.1) is the foundation of this rule.

An obvious conclusion is:

*Theorem 8*

If $R_1$ is complete with respect to $R_2$, then $R_1' \wedge (R_2(V, \underline{V}) | R_1(\underline{V}, \underline{V}))$ is $\underline{V}$_free.

## Example 15

Let $R_1 \equiv \underline{x} = x+y \wedge \underline{y} = y$ and $R_2 \equiv \underline{y} = x - y$, we have $ind(R_1, R_2) \equiv \underline{x} = x+y$, $R_1(\underline{V}, \underaccent{\tilde}{V}) \equiv \underline{x} = x+y \wedge \underline{y} = y$,

$R_2(V, \underaccent{\tilde}{V}) \equiv \underline{y} = \underaccent{\tilde}{x} - \underaccent{\tilde}{y}$ and $R_2(V, \underaccent{\tilde}{V}) \mid R_1(\underline{V}, \underaccent{\tilde}{V})$

$\equiv \underline{y} = (x+y) - y \equiv \underline{y} = x$. By Computation Rule 2:

$R_1; R_2 \rightarrow ind(R_1, R_2) \wedge (R_2(V, \underaccent{\tilde}{V}) \mid R_1(\underline{V}, \underaccent{\tilde}{V}))$,

i.e. $\underline{x} = x+y \wedge \underline{y} = y; \underline{y} = x - y \rightarrow \underline{x} = x+y \wedge \underline{y} = x$.

In the same way we can prove $\underline{x} = x+y \wedge \underline{y} = x$; $\underline{x} = x - y \rightarrow \underline{x} = y \wedge \underline{y} = x$.

This example concludes that $\underline{x} = x+y \wedge \underline{y} = y; \underline{y} = x - y;$

$\underline{x} = x - y \rightarrow \underline{x} = y \wedge \underline{y} = x$,

i.e. $\underline{x} = x+y \wedge \underline{y} = y; \underline{y} = x - y; \underline{x} = x - y$ is a specification of a program that exchanges the values of $x$ and $y$. But, to derive this conclusion formally, we need the Computation Rule below:

## Computation Rule 3: Association
$R_1; R_2; R_3 \equiv (R_1; R_2); R_3$
$R_1; R_2; R_3 \equiv R_1; (R_2; R_3)$

Theorem 2 in Section 4 is the foundation of this rule.

People would prefer to take $R' \equiv \underline{x} = x+y$ instead of $R_1 \equiv \underline{x} = x+y \wedge \underline{y} = y$ as the first SP in the specification in Example 15. But, $R_1'$ is not complete with respect to $R_2$. To prove $R_1'; R_2; R_3$ is as good as $R_1; R_2; R_3$ as a specification of exchanging values of $x$ and $y$, we need the next computation rule:

## Computation Rule 4: substitution
$R_1 \equiv R_1' \wedge R_2 \equiv R_2' \rightarrow R_1; R_2 \equiv R_1'; R_2'$

By Computation Rule 1 we know $R_1' \equiv R_1$, by Computation Rule 4 we know $R_1'; R_2 \equiv R_1; R_2$, so $R_1'; R_2; R_3 \equiv R_1; R_2; R_3$. That is, $R_1'; R_2; R_3$ is as good as $R_1; R_2; R_3$.

## Theorem 9
$R_1; R_2; R_3; R_4 \equiv (R_1; R_2); (R_3; R_4)$

## Proof

$R_1; R_2; R_3 \equiv (R_1; R_2); R_3$     (Rule 3)
$R_1; R_2; R_3; R_4 \equiv (R_1; R_2); R_3; R_4$     (Rule 4)
$(R_1; R_2); R_3; R_4 \equiv (R_1; R_2); (R_3; R_4)$     (Rule 3)

Thus, $R_1; R_2; R_3; R_4 \equiv (R_1; R_2); (R_3; R_4)$

Note that Rule 2 reduces $R_1; R_2$ to a single SP, Rule 3 allows to apply Rule 2 more than once to reduce any formula to a single $\underaccent{\tilde}{V}$_free SP. Rule 4 let us do the reduction in parallel. It is important to notice that all actual reductions always reduces two SP's to a single $\underaccent{\tilde}{V}$_free one, and the two SP's to be reduced must satisfy the requirement that "$R_1$ is complete with respect to $R_2$".

To ensure such completeness, we define the concept of reducible formulas.

## Definition 16
A SP formula $R_1; R_2; ...; R_n$, $n \geq 2$, is reducible if $R_1; R_2$ can be reduced to a single SP and $(R_1; R_2); R_3; ...; R_n$ is reducible.

With this concept, we have

## Theorem 10
1. All reducible formulas can always be reduced to a single $\underaccent{\tilde}{V}$_free SP.
2. Reducible formulas comply with associative law.

The second part of this theorem can be proved in the way Theorem 9 is proved. The first part is a direct conclusion from Definition 16.

## Example 16

$f \equiv \underline{x} \leq \underline{y}; \underline{x} \leq \underline{z}; \underline{y} \leq \underline{z}$ may be taken, by many, as a

specification of sorting $x, y, z$ into ascending order. But this formula is not complete. The OE program

$$\overline{x}(1)\overline{y}(2); \overline{x}(3)\overline{y}(4); \overline{y}(1)\overline{z}(2)$$

satisfies $f$. But this OE ends up at $\underline{x} = 3 \wedge \underline{y} = 1 \wedge \underline{z} = 2$

and $\underline{x}, \underline{y}, \underline{z}$ have nothing to do with their initial values and they are not in ascending order.

Example 18 in next section will suggest a correct specification for sorting $x, y, z$ into ascending order.

**Example 17**

For $R_1 \equiv \underline{x} = \sin x$, $R_2 \equiv \underline{x} = x^3$ and $R_3 = \sqrt{x}$, we have

$R_1; R_2 \equiv \underline{x} = (\sin x)^3$ , $(R_1; R_2); R_3 \equiv \underline{x} = \sqrt{(\sin x)^3}$ and

$R_2; R_3 \equiv \underline{x} = \sqrt{x^3}$, $R_1; (R_2; R_3) \equiv \underline{x} = \sqrt{(\sin x)^3}$.

### 7. SP CALCULUS AND SPECIFICATION ANALYSIS

Examples are used in this section to illustrate how to do specification analysis with SP calculus.

Examples 15-17 given in section 6 have in fact shown the function of SP calculus in specification analysis as long as all those formulas are considered as program specifications.

Specification process includes a phase for refinements. $\underline{x} = y \wedge \underline{y} = x$ can be refined to

$\underline{x} = x + y; \underline{y} = x - y; \underline{x} = x - y$ . The former indicates "WHAT" is wanted and the latter tells "HOW" to do it.

**Example 18**

Let be $R \equiv \{\underline{x}, \underline{y}, \underline{z}\} = \{x, y, z\}$ , $R_1 \equiv \underline{x} \leq \underline{y} \wedge \underline{z} = z$ ,

$R_2 \equiv \underline{x} \leq \underline{z} \wedge \underline{y} = y$ and $R_3 \equiv \underline{y} \leq \underline{z} \wedge \underline{x} = x$, we first prove that $f' \equiv R_1 \wedge R; R_2 \wedge R$ is a complete SP formula.
Since $R_1(\underline{V}, \underline{V}) \wedge R(\underline{V}, \underline{V}) \equiv$

$\underline{x} \leq \underline{y} \wedge \underline{z} = z \wedge \{\underline{x}, \underline{y}, \underline{z}\} = \{x, y, z\}$

and $R_2(V, \underline{V}) \wedge R(V, \underline{V}) \equiv$

$\underline{x} \leq \underline{z} \wedge \underline{y} = y \wedge \{\underline{x}, \underline{y}, \underline{z}\} = \{\underline{x}, \underline{y}, \underline{z}\}$ ,

by Rule 2:

$f' \rightarrow \underline{x} \leq \underline{z} \wedge \underline{y} = y \wedge \{\underline{x}, \underline{y}, \underline{z}\} = \{\underline{x}, \underline{y}, \underline{z}\}$

$| \underline{x} \leq y \wedge \underline{z} = z \wedge \{\underline{x}, \underline{y}, \underline{z}\} = \{x, y, z\}$

$\equiv \underline{x} \leq \underline{z} \wedge \{\underline{x}, \underline{y}, \underline{z}\} = \{x, y, z\} \wedge \underline{x} \leq y$.

where $\underline{x} \leq y$ is obtained in the following way:

$\underline{x} \leq \underline{z} \wedge \underline{y} = y \wedge \{\underline{x}, \underline{y}, \underline{z}\} = \{\underline{x}, \underline{y}, \underline{z}\}$

$\rightarrow \underline{x} \leq \underline{z} \wedge \{\underline{x}, \underline{z}\} = \{\underline{x}, \underline{z}\}$

$\rightarrow \underline{x} \leq \underline{x}$, since $\underline{x}$ is the smaller one in the set.

But, $\underline{x} \leq \underline{y}$ and $\underline{y} = y$ , so $\underline{x} \leq \underline{y}$ . Thus $\underline{x} \leq \underline{x}$ leads to $\underline{x} \leq y$ .

That is, $f'$ is complete, and

$f' \rightarrow \underline{x} \leq y \wedge \underline{x} \leq \underline{z} \wedge R$.

Similarly we can prove that

$R_1 \wedge R; R_2 \wedge R; R_3 \wedge R \rightarrow \underline{x} \leq y \wedge \underline{y} \leq \underline{z} \wedge R$.

This is a proper specification of a sorting program.

**Example 19**

$f(n) \equiv \underline{i} = 1 \wedge \underline{f} = 1; (\underline{i} = i + 1 \wedge \underline{f} = f \cdot i)^n$

is a specification for the program to compute $n!$ for $n \geq 1$,
since

$f(n) \rightarrow \underline{f} = n! \wedge \underline{i} = n + 1$

If we take the Boolean constant True as a SP, and for $R$, $R$ is an arbitrary SP, define $R^0 =$ True, then $f(n)$ can be relaxed to allow $n = 0$. We have, by Rule

2:

$$f(0) \to \underline{i} = 1 \wedge \underline{f} = 1 \quad \text{since } \underline{i} = 1 \wedge \underline{f} = 1 \text{ is}$$

independent of True. i.e. $f(0) \to \underline{f} = 0!$

The prove of $f(n) \to \underline{f} = n! \wedge \underline{i} = n+1$ requires the application of mathematical induction.

**Example 20**

A specification in terms of SP calculus will be proposed for the 8_Queen Problem.

The first step towards a formalization of this problem is to do abstraction: to find a mathematical representation of physical objects involved, namely the chess board and the queens on the board.

To keep data structure as simple as possible, we use an array as the abstraction of the board, i.e. array $A[0..7]$ whose elements take values from $\{0,1,2,3,4,5,6,7\}$, and $A[i] = j$ denotes that a queen is placed at $(i, j)$ on the board.

The second step is to represent in terms of $A[0..7]$ what the problem is required: to place 8 queens on board such that they stay in peace, i.e. no one is in a position to kill anyone else. For two queens $A[i] = l$ and $A[j] = k$, they can not kill each other if and only if

$$l \neq k \wedge l - k \neq i - j \wedge l - k \neq j - i$$

i.e. $A[i] \neq A[j] \wedge A[i] - A[j] \neq i - j \wedge A[i] - A[j] \neq j - i$.

We denote this with $peace(A, i, j)$, i.e.

$peace(A, i, j)$
$\equiv A[i] \neq A[j] \wedge A[i] - A[j] \neq i - j \wedge A[i] - A[j] \neq j - i$

The third step is to propose a solution.

There are two ways to understand this problem: to find one solution for the 8 queens, or to find all solutions for the 8 queens. The auxiliary variables proposed next aim at all solutions.

Auxiliary variables:

$i, l$: integer with $-1 \leq i \leq 7$ and $0 \leq l \leq 8$.

$P$: $array[0..7]$ of integer whose element set is $\{0,1,2,3,4,5,6,7\}$.

$Q$: constant $array[0..7]$ with all elements being 0.

intended meaning:

$l < 9 \wedge i \geq 0 \to A[i] = l$ is a candidate for the $(i+1)th$ queen.

$l = 8 \to A[0..i-1]$ is not part of next solution.

$i = -1 \to$ no more solution

$i \geq 0 \to A[0..i-1]$ may be part of next solution.

We need two predicates:

$R(A, i) \equiv \forall j, k : 0 \leq j, k < i :: peace(A, j, k)$, for $i, 0 \leq i \leq 8$

$H(A, i, l) \equiv (\underline{A}[i] = l \to R(A, i+1))$

We have

$R(A, 8) \to (A \text{ is a solution})$

Specification for finding one solution:

Let be

$R_0 \equiv \underline{i} = 1 \wedge \underline{l} = 0 \wedge \underline{P} = Q \wedge \underline{A} = Q$

$R_1 \equiv \underline{l} = 8 \vee (\underline{l} < 8 \wedge H(A, i, \underline{l}))$

$R_2 \equiv l < 8 \to \underline{A}[i] = l$

$R_3 \equiv (i = 7 \wedge l < 8 \to \underline{P} = A)$
$\quad \oplus (i < 7 \wedge l < 8 \to \underline{i} = i+1 \wedge \underline{l} = 0)$
$\quad \oplus (i > 0 \wedge l = 8 \to \underline{i} = i-1 \wedge \underline{l} = A[i-1]+1)$
$\quad \oplus (l = 8 \wedge i = 0 \to \underline{i} = -1)$

The solution is

$R_0 ; (R_1; R_2; R_3)^{\underline{P} \neq Q \vee \underline{i} = -1}$

This solution includes a formula $(R_1; R_2; R_3)^{\underline{P} \neq Q \vee \underline{i} = -1}$, and this formula can not be derived from the definition of SP_formulas. We may extend the definition to include this, but we didn't. We have treated $R^n$ as an abbreviation of $R; R; \cdots; R$ for simplicity. $R^b$, where $b$ is a SP, can also be viewed as an abbreviation of $R; R; \cdots; R$, with a flexible times of repetition of $R$. If this repetition does not lead to $\underline{b}$, $R^b$ is not a right formula. For one solution, $\underline{P} \neq Q \vee \underline{i} = -1$ can be reached, since when $A$ is considered as an octal number, $A < \underline{A}$ is derivable from $(R_1; R_2; R_3)^n$ for some $n$ unless $\underline{P} \neq Q \vee \underline{i} = -1$.

$\underline{P} \neq Q$ denotes the fact that $P$ is a solution, and $\underline{i} = -1$ means that there is no solution.

Specification for finding all solutions:
Let be
$$R_3' \equiv (i = 7 \wedge l < 8 \rightarrow \underline{P} = A \wedge \underline{l} = A[6]+1 \wedge \underline{i} = 6)$$
$$\oplus (i < 7 \wedge l < 8 \rightarrow \underline{i} = i+1 \wedge \underline{l} = 0)$$
$$\oplus (i > 0 \wedge l = 8 \rightarrow \underline{i} = i-1 \wedge \underline{l} = A[i-1]+1)$$
$$\oplus (l = 8 \wedge i = 0 \rightarrow \underline{i} = -1)$$

The solution is

$$R_0; (R_1; R_2; R_3')^{i=-1}$$

and each time $P$ is changed, a new solution is found.

Note that the difference between a specification and a program is that a program consists of details of operations while a specification contains only the result of these operations. For example, $R_1$ demands $\underline{l}$ to be either $\underline{l} = 8$, or $\underline{l} < 8 \wedge H(A, i, \underline{l})$, it does not tell how to achieve the goal.

## 8. HOW TO DEVELOP SOE FROM GIVEN SPECIFICATIONS

The SOE for $\underline{x} = y \wedge \underline{y} = x$ is $\bar{x}(y)\bar{y}(x)$

The SOE for $\underline{x} = x+y; \underline{y} = x-y; \underline{x} = x-y$ is

$\bar{x}(x+y); \bar{y}(x-y); \bar{x}(x-y)$

The SOE for $\underline{x} \leq \underline{y} \wedge R; \underline{x} \leq \underline{z} \wedge R; \underline{y} \leq \underline{z} \wedge R$, where

$R \equiv \{\underline{x}, \underline{y}, \underline{z}\} = \{x, y, z\}$, is

$(\bar{x}(y)\bar{y}(x))^{x>y}; (\bar{x}(z)\bar{z}(x))^{x>z}; (\bar{y}(z)\bar{z}(y))^{y>z}$

The SOE for $\underline{i} = 1 \wedge \underline{f} = 1; (\underline{i} = i+1 \wedge \underline{f} = f*i)^n$ is

$\bar{i}(1)\bar{f}(1); (\bar{i}(i+1)\bar{f}(f*i))^n$.

The SOE for the 8_Queen problem is:

SOE for $R_0 : P_0 \equiv \bar{i}(1)\bar{l}(0)\bar{P}(Q)\bar{A}(Q)$

SOE for $R_1 : P_1 \equiv (\bar{l}(l+1))^{l=8 \vee H(A,i,l)}$

SOE for $R_2 : P_2 \equiv \bar{A}[i](l)^{l<8}$

SOE for $R_3$:

$P_4 \equiv \bar{P}(A)^{i=7 \wedge l=8} \cdot (\bar{i}(i+1)\bar{l}(0))^{i<7 \wedge l=8} \cdot$

$(\bar{i}(i-1)\bar{l}(A[i-1]+1))^{i>0 \wedge l=8} \cdot \bar{i}(-1)^{l=8 \wedge i=0}$

SOE for $R_3'$:

$P_4 \equiv (\bar{P}(A) \cdot \bar{i}(6)\bar{l}(A[6]+1))^{i=7 \wedge l=8} \cdot$

$(\bar{i}(i+1)\bar{l}(0))^{i<7 \wedge l=8} \cdot (\bar{i}(i-1)\bar{l}(A[i-1]+1))^{i>0 \wedge l=8} \cdot (\bar{i}(-1))^{l=8 \wedge i=0}$

SOE for one solution $P_0; (P_1; P_2; P_3)^{P \neq Q \vee i = -1}$

SOE for all solutions $P_0; (P_1; P_2; P_4)^{i=-1}$

Apparently, it is not difficult to develop a SOE from a given SP_formula since for $R$ ($R$ is a SP), $V_R$ is the set of variables that require a simultaneous write operation.

The analysis of the specifications for the 8_Queen problem has been omitted here to save space. In fact, this problem is among those that are used for demonstration of prototype tools developed in our lab for SOE analysis.

The prototype tools are in fact a SOE property verifier capable to do semi_automatic SOE analysis (human interfere is needed when, say, mathematical induction is required). A SOE may be re_structured as a tree, called analysis tree, whose intermediate nodes are control operators for operation sequences, choices and loops, while leaf nodes are SOE Terms. We go no further here on these tools since this paper focuses on OESPA, the theory itself.

Advantages of SOE and SP calculus include:
- SOE is syntactically formal as well as semantically formal. As such, properties can be

computed from SOE texts.

- SP, i.e. semantic predicates, and SP formulas provide a firm foundation for program specification, since direct descriptions of how consecutive program states are related become feasible. SP calculus has made specification analysis practical.
- Specification in SP formulas let users focus on WHAT instead of HOW, since operation details are left for later consideration.

To end this section, we have the following definition:

**Definition 17**

A SOE developed from a given reducible SP formula is correct if and only if the given formula is a property of it.

Note that, in our theory OESPA, a SOE is a program, a reducible SP formula is a program specification, and a reducible SP formula can always be reduced to a single SP.

## 9. SOE And Software Engineering

The software industry depends heavily on software engineering. But software engineering relies on post-development testing for the quality of its products.

It is impossible, at least for the time being, to have SOE as the main means of programming. But SOE and SP_calculus provide a formal foundation for software engineering. With the help of SOE and SP calculus, programmers may better understand what programming is about.

We wish to have a programming language that has a BNF definition similar to programming languages in wide use, but potentially equivalent to the BNF definition of SOE. In this way, software qualities may be guaranteed with testing being of second importance only.

We have applied our theory to develop tools for the analysis of C pointers. It has turned out to be very helpful. It leads us to a systematical way of thinking, and makes it possible to develop a C-pointer analysis prototype tool within half a year.

## 10. Future Work

This paper focused on SOE, sequential operation expression. We will have POE and ROE proposed in the future: POE is an abbreviation for Parallel Operation Expression while ROE, an abbreviation for Reactive Operation Expression.

Parallelism is the only means to promote software efficiency and to make full use of super computers. Reactive systems are important for computer applications in the area of management and control.

## Acknowledgments

We are grateful to the National Engineering Research Center For Software Engineering of Peking University, Beijing Beida Software Engineering Development Company for their continuous support. Thanks to everyone in our lab among them Mr. Zhang Hang has helped us with typesetting. The authors are sponsored by the National Basic Research Program of China (973) No.2009CB320706.

## References

[1] Emerson Clarke, Formal Methods: State of the Art and Future Directions [R]. VeriSign Inc, Mountain View, CA, 2004.

[2] E.M. Clarke, O. Grumberg, and D.A. Peled, Model Checking[C]. MIT Press, 1999.

[3] Hoare C A R. Communicating sequential processes[M]. Prentice Hall，1985.

[4] Hoare C A R. An axiomatic basis for computer programming[J]. Communications of the ACM, 1969, 12(10): 576-580.

[5] C.A.R.Hoare, HeJifeng. Unifying Theories of Programming[M].Prentice Hall International，1998

[6] Chongyi Yuan, Assignment：Operation on a

physical object[J]. Journal of Frontiers of Computer Science and Technology，2008，2（5）：487-499.

[7] Chongyi Yuan, Huang Yu, Zhao Wen. 2009. Program:Expressions of Operations On Physical Objects. Journal of Frontiers of Computer Science and Technology. 2009 Vol.3(2):144-153.

[8] Chongyi Yuan, Zhao Wen, Gao Xin, Huang Yu. 2010. Definitions and Specifications of O-expression properties. Journal of Frontiers of Computer Science and Technology. 2010 Vol.4(1):20-28.

[9] Chongyi Yuan, Wen Zhao, Yu Huang. Operation Expression: A Way to Verified Software. The 2010 International Conference on Foundations of Computer Science (FCS2010, WORLDCOMP'10), Las Vegas Nevada, USA, 2010, 91-96.

[10] Dijkstra E W. Letters to the editor: go to statement considered harmful[J]. Communications of the ACM, 1968, 11(3): 147-148.